\documentclass[12pt]{article}
\usepackage{amsmath, amssymb, amsthm}
\usepackage{booktabs}
\usepackage{graphicx}
\usepackage{hyperref}
\usepackage[margin=1in]{geometry}
\usepackage{natbib}
\usepackage{algorithm}
\usepackage{algpseudocode}
\usepackage{xcolor}
\usepackage{enumitem}

\newcommand{\bY}{\mathbf{Y}}
\newcommand{\bX}{\mathbf{X}}
\newcommand{\bZ}{\mathbf{Z}}
\newcommand{\bbeta}{\boldsymbol{\beta}}
\newcommand{\balpha}{\boldsymbol{\alpha}}
\newcommand{\bgamma}{\boldsymbol{\gamma}}

\newcommand{\bSigma}{\boldsymbol{\Sigma}}
\newcommand{\bD}{\mathbf{D}}
\newcommand{\bI}{\mathbf{I}}
\newcommand{\bZG}{\mathbf{Z}_\Gamma}
\newcommand{\bone}{\mathbf{1}}
\newcommand{\bu}{\mathbf{u}}
\newcommand{\bvarepsilon}{\boldsymbol{\varepsilon}}
\newcommand{\Norm}{\mathcal{N}}
\newcommand{\tauconv}{\tau_{\mathrm{conv}}}

\title{SEMMS with Random Effects:\\
       A Mixed-Model Extension for Variable Selection\\
       in Clustered and Longitudinal Data}
\author{Haim Bar \and Martin T. Wells}
\date{\today}

\begin{document}
\maketitle

\begin{abstract}
SEMMS (Scalable Empirical-Bayes Model for Marker Selection) is a variable-selection
procedure for generalized linear models that uses a three-component normal mixture
prior on regression coefficients.  In its original form, SEMMS assumes that all observations
are independent. Many real-world datasets, however, arise from repeated-measures or
clustered designs in which observations within the same subject are correlated.
Ignoring this correlation inflates the apparent residual variance and can severely
degrade variable-selection performance.
We extend SEMMS to accommodate random intercepts, random slopes, or both, via an
alternating coordinate-ascent algorithm.  After each round of fixed-effect variable
selection, the subject-level best linear unbiased predictors (BLUPs) are updated with
\texttt{lmer} (Gaussian) or \texttt{glmer} (non-Gaussian); the fixed-effect step then
operates on the random-effect-adjusted response.
We describe the algorithm, evaluate its performance in three Gaussian simulation studies
spanning a range of signal strengths, random-effect magnitudes, and sample/predictor-space
regimes, and present a semi-synthetic real-data example.  We further extend the framework to non-Gaussian families (Poisson, binomial)
via an IRLS working-response adaptation \citep{breslow1993}: at each outer
iteration the fixed-effects step uses the RE-adjusted working response computed
from the current \texttt{glmer} fitted values rather than the raw response.
When the fixed-effect signal is strong relative to the random-effect variance,
both the original and extended procedures perform comparably.  When the
random-effect variance dominates---the scenario most likely to cause plain
SEMMS to fail---the mixed-model extension recovers the exact true predictor
set in 93\% of simulated datasets (Gaussian), 61\% (Poisson), and 65\%
(binomial), compared with 1\%, 45\%, and 39\% for plain SEMMS respectively.
\end{abstract}

%% ============================================================
\section{The SEMMS Model}
\label{sec:semms}
%% ============================================================

\subsection{Setup and notation}

Let $\bY = (Y_1,\ldots,Y_N)^\top$ be a continuous response and
$\bZ = [\mathbf{z}_1\,\cdots\,\mathbf{z}_K]$ an $N\times K$ matrix of
standardized candidate predictors.  A fixed-effect design matrix
$\bX\in\mathbb{R}^{N\times P}$ may include an intercept and any covariates
that are to be retained regardless of the variable-selection outcome.
SEMMS was introduced by \citet{bar2019} for variable selection in
generalized linear models using an empirical Bayes framework.

\subsection{The mixture prior}

Each predictor $k = 1,\ldots,K$ is assigned a latent state
$\gamma_k\in\{-1,0,+1\}$, where $0$ denotes a null effect, $+1$ a positive
effect and $-1$ a negative effect.  Writing $\mathcal{S} = \{k:\gamma_k\ne 0\}$
for the active set with $|\mathcal{S}|=L$, define
\[
  \bZG = \bZ_{\mathcal{S}}\,\mathrm{diag}(\bgamma_{\mathcal{S}}),
  \qquad L_- = |\{k:\gamma_k=-1\}|,\quad
  L_+ = |\{k:\gamma_k=+1\}|,\quad
  L_0 = K-L.
\]
The working model for the Gaussian case is
\[
  \bY = \bX\bbeta + \bZG\,\mu\bone_L + \bvarepsilon,
  \qquad
  \bvarepsilon \sim \Norm\!\bigl(\mathbf{0},\;\bSigma\bigr),
\]
where $\mu>0$ is the magnitude of the common effect shared by all active predictors,
$\bone_L$ is an $L$-vector of ones, and
\begin{equation}
  \bSigma = \sigma^2_e\,\bI_N + \sigma^2_r\,\bZG\bZG^\top.
  \label{eq:sigma}
\end{equation}
The covariance structure~\eqref{eq:sigma} arises from a hierarchical
representation in which the true coefficient of each selected variable is
$\mu\gamma_k + \eta_k$ with $\eta_k\overset{\mathrm{iid}}{\sim}\Norm(0,\sigma^2_r)$,
so that the active coefficients share a common signed mean $\pm\mu$ but are
allowed to deviate from it.  Here $\sigma^2_e$ is the pure observation noise.

A log-prior over the component counts contributes
\[
  \ell_{\mathrm{prior}}(\bgamma) =
  \sum_{s\in\{-,0,+\}} L_s \log\!\Bigl(\tfrac{L_s}{K}\Bigr)
\]
to the objective, acting as a multiplicity penalty that discourages
unnecessarily large active sets.

\subsection{Log-likelihood via the Woodbury identity}

Direct evaluation of $\log|\bSigma|$ and the quadratic form
$\bvarepsilon^\top\bSigma^{-1}\bvarepsilon$ would require $O(N^3)$ operations
for arbitrary $N$.  SEMMS~v0.2.5 exploits the matrix determinant lemma and the
Woodbury matrix identity to reduce all computations to $O(NL^2)$:
\begin{align}
  \bSigma^{-1} &= \tfrac{1}{\sigma^2_e}\bI_N
    - \tfrac{\sigma^2_r}{\sigma^4_e}\bZG\,\mathbf{B}\,\bZG^\top,
    \qquad
    \mathbf{B} = \Bigl(\bI_L + \tfrac{\sigma^2_r}{\sigma^2_e}\bZG^\top\bZG\Bigr)^{-1},
  \label{eq:woodbury} \\
  \log|\bSigma^{-1}| &= -N\log\sigma^2_e
    - \log\Bigl|\bI_L + \tfrac{\sigma^2_r}{\sigma^2_e}\bZG^\top\bZG\Bigr|.
\end{align}
Since $L\ll N$ in practice, the $L\times L$ matrix $\mathbf{B}$ is cheap to
form and invert, and no $N\times N$ matrix ever materializes.

\subsection{The GAM algorithm}

SEMMS fits the model via a Generalized Alternating Maximization (GAM) algorithm
\citep{gunawardana2005} that cycles through:
\begin{enumerate}[leftmargin=*, label=\arabic*.]
  \item \textbf{Variable selection step.}  For each candidate predictor $k$
    and each state $s\in\{-1,0,+1\}$, compute the change in
    $\ell_{\mathrm{prior}} + \ell_{\mathrm{lik}}$ that would result from
    reassigning $\gamma_k\leftarrow s$, holding all other assignments fixed.
    In the greedy variant, the $(k^*,s^*)$ pair with the largest gain
    $>\Delta_{\min}$ is accepted.
  \item \textbf{Parameter update (inner EM).}  Given the current active set,
    run the EM algorithm to update $(\mu,\bbeta,\sigma^2_e,\sigma^2_r)$.
    Pre-computed quantities $\bZG^\top\bZG$, $\mathbf{H}^\top\mathbf{H}$,
    $\mathbf{H}^\top\bY$ (where $\mathbf{H}=[\bX\mid\bZG]$) remain fixed
    for the life of the inner loop; only the $L\times L$ matrix $\mathbf{B}$
    needs updating at each EM step as $\sigma^2_e$ and $\sigma^2_r$ change.
  \item Repeat until the log-likelihood increase falls below $\Delta_{\min}$
    or a maximum iteration count is reached.
\end{enumerate}
The algorithm is guaranteed to produce a non-decreasing sequence of
log-likelihood values.  For non-Gaussian families (Poisson, binomial), SEMMS
uses an iteratively reweighted least squares (IRLS) linearization, updating a
working response $\tilde{\bY}$ at each outer iteration.

%% ============================================================
\section{Random-Effects Extension}
\label{sec:mixed}
%% ============================================================

\subsection{Motivation and use cases}

The independence assumption underlying plain SEMMS is violated in several
common study designs.

\begin{itemize}
  \item \textbf{Longitudinal studies.} Biomarker measurements
 of a patient collected at multiple time points share a subject-specific baseline (random
    intercept) and possibly a subject-specific trajectory (random slope in time).
    Failing to account for within-subject correlation inflates the apparent
    residual variance, making it harder to detect true predictors and
    easier to select noise.

  \item \textbf{Repeated-measures experiments.}  In pharmacology or psychology,
    each participant receives multiple treatments or stimuli.  Within-subject
    correlation is the rule, not the exception.

  \item \textbf{Clustered observational data.}  Patients nested within hospitals,
    students nested within schools, or specimens nested within batches all
    exhibit intra-cluster correlation that should be modeled explicitly.

  \item \textbf{Twin or family studies.} Genetic relatedness induces
    correlations between observations that can confound naive variable-selection
    procedures.

  \item \textbf{Multi-site genomic studies.}  Expression data collected across
    multiple laboratories or cohorts often show site-level batch effects
    that act as random intercepts.
\end{itemize}

In all these settings, the within-subject correlation structure is a nuisance
that must be removed before the fixed-effect signal can be reliably detected.

\subsection{The extended model}

For a longitudinal design with $m$ subjects and $n_i$ observations per subject,
index observations by $(i,j)$ where $i=1,\ldots,m$ and $j=1,\ldots,n_i$.
The extended model is
\begin{equation}
  Y_{ij} = \mathbf{x}_{ij}^\top\bbeta
          + \mathbf{z}_{ij}^\top\balpha
          + b_{0i} + b_{1i}\,t_j
          + \varepsilon_{ij},
  \label{eq:mixed}
\end{equation}
where
\begin{itemize}
  \item $\bbeta\in\mathbb{R}^P$ are fixed effects for covariates in $\bX$,
  \item $\balpha\in\mathbb{R}^K$ are the candidate-predictor coefficients,
    to be selected and estimated by SEMMS via the mixture model,
  \item $b_{0i}$ is a subject-specific random intercept,
  \item $b_{1i}$ is a subject-specific random slope on a continuous variable
    $t$ (e.g.\ time, dose),
  \item $\varepsilon_{ij}\overset{\mathrm{iid}}{\sim}\Norm(0,\sigma^2_e)$,
\end{itemize}
and the random effects satisfy
\[
  \begin{pmatrix}b_{0i}\\b_{1i}\end{pmatrix}
  \overset{\mathrm{iid}}{\sim}
  \Norm\!\left(\mathbf{0},\;
    \bD = \begin{pmatrix}\sigma^2_{b_0} & \rho\sigma_{b_0}\sigma_{b_1}\\
                          \rho\sigma_{b_0}\sigma_{b_1} & \sigma^2_{b_1}\end{pmatrix}
  \right).
\]
A user may include only a random intercept (by setting $\sigma^2_{b_1}=0$),
only a random slope (setting $\sigma^2_{b_0}=0$), or both.

Let $\hat{\bu}_{ij} = \hat{b}_{0i} + \hat{b}_{1i}t_j$ denote the
vector of BLUP contributions from the random effects.  The key insight
underlying the algorithm is that, \emph{conditional on $\hat{\bu}$}, the
adjusted response is
\[
  Y^*_{ij} = Y_{ij} - \hat{u}_{ij}
\]
satisfies the standard SEMMS independence assumption, so that the
existing GAM algorithm \citep{gunawardana2005} can be applied without modification. In contrast,
\emph{conditional on the currently selected fixed effects}, the random-effect
parameters $(\bD,\sigma^2_e)$ can be consistently estimated by maximum
likelihood via \texttt{lmer} \citep{bates2015}.

This decoupling motivates a coordinate-ascent outer loop that alternates
between the two conditional steps.

\subsection{Algorithm}

\begin{algorithm}[H]
\caption{\textsc{fitSEMMSmixed}: alternating fixed/random-effects updates}
\label{alg:mixed}
\begin{algorithmic}[1]
\Require Data $(\bY,\bX,\bZ,\text{group},t)$;
         RE specification (intercept, slope, or both);
         SEMMS hyperparameters; outer tolerance $\tauconv$;
         maximum outer iterations $T$.
\Ensure Selected active set $\hat{\mathcal{S}}$; fitted REML \texttt{lmer}.

\State $\hat{\bu}^{(0)} \leftarrow \mathbf{0}$
\Comment{initialize RE offset at zero}

\State Fit null \texttt{lmer} (intercept-only fixed part) with ML;
       update $\hat{\bu}^{(0)} \leftarrow
       \hat{\bY}_{\text{full}} - \hat{\bY}_{\text{fixed}}$
\Comment{warm-start from subject baselines}

\For{$t = 1, 2, \ldots, T$}
  \State $\bY^* \leftarrow \bY - \hat{\bu}^{(t-1)}$
  \Comment{partial out current RE contribution}

  \State Run \textsc{fitSEMMS}$(\bY^*, \bX, \bZ, \ldots)$;
         obtain active set $\hat{\mathcal{S}}^{(t)}$
  \Comment{fixed-effects step; C++ GAMupdate unchanged}

  \State Fit \texttt{lmer}$(\bY \sim \bX + \bZ_{\hat{\mathcal{S}}^{(t)}}
         + \text{RE formula},\;\text{REML}=\text{FALSE})$
  \Comment{random-effects step (ML for alternating loop)}

  \State $\hat{\bu}^{(t)} \leftarrow
         \hat{\bY}_{\text{full}}^{(t)} - \hat{\bY}_{\text{fixed}}^{(t)}$
  \Comment{new BLUP vector, length $N$}

  \If{$\max_i\bigl|\hat{u}^{(t)}_i - \hat{u}^{(t-1)}_i\bigr| < \tauconv$}
    \State \textbf{break}
    \Comment{outer loop converged}
  \EndIf
\EndFor

\State Refit \texttt{lmer} with $\hat{\mathcal{S}}^{(t)}$ and $\text{REML}=\text{TRUE}$
\Comment{final inference}

\State \Return $\hat{\mathcal{S}}^{(t)}$, REML \texttt{lmer}, $\hat{\bu}^{(t)}$
\end{algorithmic}
\end{algorithm}

\paragraph{Warm-start rationale.}
Initializing $\hat{\bu}^{(0)}$ from a null \texttt{lmer} (rather than from
zero) is important when subject-level random effects are large relative to
the fixed-effect signal.  Without warm-starting, SEMMS may absorb
subject-level variation into selected predictors during the first iteration,
potentially locking in false positives that are difficult to remove in
subsequent iterations.

\paragraph{Convergence guarantee.}
Each outer iteration leaves the joint log-likelihood non-decreasing: the
fixed-effects step maximizes over $\hat{\mathcal{S}}$ given $\hat{\bu}$,
and the RE step maximizes over the random-effect parameters given
$\hat{\mathcal{S}}$.  This coordinate-ascent property guarantees convergence
to a local optimum of the joint log-likelihood.

\paragraph{ML vs.\ REML.}
\texttt{lmer} is called with \texttt{REML = FALSE} during the alternating loop
so that likelihoods are comparable across iterations with different fixed-effect
sets.  The final call uses \texttt{REML = TRUE} to obtain unbiased variance
component estimates for reporting.

\subsection{Implementation}

The extension is implemented as two new functions in the \texttt{SEMMS} R
package (v0.4.0) \citep{SEMMS-R}.

\begin{itemize}
  \item \texttt{readInputFile()} gains two optional parameters:
    \texttt{group\_col} (column index of the subject/cluster identifier) and
    \texttt{random\_slope\_col} (column index of the random-slope covariate,
    e.g.\ time).  These columns are stored in the returned \texttt{dat} list as
    \texttt{dat\$group} and \texttt{dat\$random\_slope} and are never included
    in the candidate predictor matrix~$\bZ$.

  \item \texttt{fitSEMMSmixed(dat, random\_intercept, random\_slope, \ldots)}
    implements Algorithm~\ref{alg:mixed}.  All SEMMS hyperparameters
    (\texttt{nn}, \texttt{mincor}, \texttt{minchange}, etc.) are forwarded
    verbatim to the inner \textsc{fitSEMMS} call.

  \item \texttt{runMixedModel(dat, nnt, re\_term)} mirrors the interface of
    \texttt{runLinearModel()}, refitting the final REML \texttt{lmer} and
    returning the model object, AIC, and variance inflation factors.
\end{itemize}

The C++ core (\texttt{GAMupdate}, \texttt{fit\_gam}, \texttt{loglik}) is
entirely unchanged.  The Woodbury-based computational gains of SEMMS~v0.2.5
carry over without modification, and the fixed-effect step remains $O(NL^2)$
per EM step.

A typical workflow is:
\begin{verbatim}
dat <- readInputFile(fn, ycol = 1, Zcols = 4:103,
                     group_col = 2, random_slope_col = 3)

fit <- fitSEMMSmixed(dat,
                     random_intercept = TRUE,
                     random_slope     = TRUE,
                     nn = 5, verbose = TRUE)

mm  <- runMixedModel(dat, fit$gam.out$nn, fit$re_term)
summary(mm$mod)
\end{verbatim}

%% ============================================================
\section{Simulation Studies}
\label{sec:sim}
%% ============================================================

We conducted three simulation studies spanning contrasting signal-to-noise
ratios, random-effect magnitudes, and sample-size/dimensionality regimes,
in order to examine the performance of the mixed-model extension relative
to plain SEMMS across a practically relevant range of scenarios.

\subsection{Common design}
\label{sec:simdesign}

All datasets were generated according to model~\eqref{eq:mixed} with:
\begin{itemize}
  \item $m = 20$ subjects, each observed at $n = 10$ equally spaced time
    points, giving $N = 200$ total observations;
  \item $K = 100$ candidate predictors drawn i.i.d.\ from $\Norm(0,1)$, of
    which $K_{\text{true}} = 5$ are truly associated with the response
    (predictors $V_1,\ldots,V_5$);
  \item the random-slope covariate $t_j\in\{1,\ldots,10\}$ is standardized
    to zero mean and unit variance before simulation;
  \item $\varepsilon_{ij}\overset{\mathrm{iid}}{\sim}\Norm(0,1)$.
\end{itemize}
The fixed-effect design matrix $\bX$ contains only an intercept.
Each simulation study consists of 100 independent replications with a fixed
random seed for reproducibility.

\paragraph{Competing methods.}
For each replication, we fit four methods:

\begin{enumerate}
  \item \textbf{Plain SEMMS}: \texttt{fitSEMMS(dat, distribution="N",
    nn=5, mincor=0.7)}.  Treats all observations as independent; does not
    account for random effects.

  \item \textbf{Mixed SEMMS}: \texttt{fitSEMMSmixed(dat,
    random\_intercept=TRUE, random\_slope=TRUE, nn=5, mincor=0.7)}.
    The proposed extension with alternating fixed/random-effect updates.

  \item \textbf{Plain lasso}: \texttt{cv.glmnet} \citep{glmnet2010} with
    5-fold cross-validation and the \texttt{lambda.min} rule applied to the
    raw response.  Like plain SEMMS, this method ignores the random-effects
    structure.

  \item \textbf{glmmLasso}: \texttt{glmmLasso} \citep{glmmlasso2014} with a
    random intercept and random slope on \texttt{time}.  The penalty
    parameter $\lambda$ is selected by BIC over a grid of 12 log-spaced
    values from 300 to 30; each grid point is fitted independently (no
    warm-starting) to avoid local-minima issues.
\end{enumerate}

We report mean true positives (TP), mean false positives (FP) and
\emph{exact recovery rate}: the proportion of replications in which the
selected set is exactly equal to $\{V_1,\ldots,V_5\}$.

\subsection{Simulation~1: moderate random effects, strong signal}
\label{sec:sim1}

\paragraph{Parameters.}
$\bbeta_{\text{true}} = (1.5,-1.2,1.0,-0.9,0.8)^\top$,
$\sigma_{b_0} = 1.5$, $\sigma_{b_1} = 0.5$.
The oracle fixed-effects-only $R^2$ is approximately $0.55$; the random
effects account for roughly one third of total variance.

\paragraph{Results.}
Table~\ref{tab:sim1} summarizes the selection performance of all four methods.
Both SEMMS variants achieve near-perfect true-positive recovery.
Mixed SEMMS achieves 100\% exact recovery with zero mean false positives;
plain SEMMS is close behind with 94\% exact recovery, confirming that moderate
random effects do not severely harm the independence-based procedure when the
signal is strong.

The plain lasso finds all five true predictors in every replication but
appends a large number of false positives (mean 13.7), because without a
random-effects model the within-subject correlation inflates the apparent
signal of many noise predictors.  glmmLasso correctly accounts for the
correlation structure and eliminates most false positives (mean 2.57),
but its BIC-selected $\lambda$ tends to be insufficiently large, yielding
exact recovery in only 25\% of replications.
Mixed SEMMS converges in at most three outer iterations (mean 2.01).

\begin{table}[ht]
\centering
\caption{Simulation~1 results ($\sigma_{b_0}=1.5$, $\sigma_{b_1}=0.5$,
         strong signal; 100 replications, $K=100$, $K_{\text{true}}=5$).}
\label{tab:sim1}
\begin{tabular}{lrrr}
\toprule
Method         & Mean TP & Mean FP & Exact recovery \\
\midrule
Plain SEMMS    & 4.98    & 0.05    & 94\% \\
Mixed SEMMS    & 5.00    & 0.00    & \textbf{100\%} \\
Plain lasso    & 5.00    & 13.69   & 0\%  \\
glmmLasso      & 4.99    & 2.57    & 25\% \\
\bottomrule
\end{tabular}
\end{table}

\subsection{Simulation~2: dominant random effects, weak signal}
\label{sec:sim2}

\paragraph{Parameters.}
$\bbeta_{\text{true}} = (0.8,-0.7,0.6,-0.6,0.5)^\top$,
$\sigma_{b_0} = 3.0$, $\sigma_{b_1} = 1.0$.
The fixed effects now account for only about 9\% of total variance, while
the random effects dominate.

\paragraph{Results.}
Table~\ref{tab:sim2} tells a markedly different story.
Mixed SEMMS remains effective, recovering all five true predictors in 93 of
100 replications (93\% exact recovery) with negligible false positives
(mean 0.02). All other methods fail.

Plain SEMMS is unable to separate fixed-effect signal from the dominant
within-subject noise: mean TP falls to 2.48 and exact recovery is near zero
(1\%).  Plain lasso selects on average only 3.56 true predictors while
adding more than 9 false positives; without a random-effects correction, it
mistakes subject-level correlation structure for signal.  glmmLasso, despite
fitting a random-effects model, performs worst of all (mean TP 0.60, exact
1\%): the BIC criterion over-penalizes the fixed effects when
$\sigma^2_{b_0}$ is large, because the random intercept absorbs most of the
response variance and leaves too little residual signal for the lasso
component to detect.

\begin{table}[ht]
\centering
\caption{Simulation~2 results ($\sigma_{b_0}=3.0$, $\sigma_{b_1}=1.0$,
         weak signal; 100 replications, $K=100$, $K_{\text{true}}=5$).}
\label{tab:sim2}
\begin{tabular}{lrrr}
\toprule
Method         & Mean TP & Mean FP & Exact recovery \\
\midrule
Plain SEMMS    & 2.48    & 1.14    & 1\%  \\
Mixed SEMMS    & 4.95    & 0.02    & \textbf{93\%} \\
Plain lasso    & 3.56    & 9.25    & 0\%  \\
glmmLasso      & 0.60    & 0.16    & 1\%  \\
\bottomrule
\end{tabular}
\end{table}

\paragraph{Discussion.}
The contrast between the two scenarios isolates the mechanism that the
mixed-model extension addresses.  In Simulation~2, the between-subject
variance $\sigma^2_{b_0}=9$ dwarfs the total fixed-effect variance
($\sum_k \beta_k^2 \approx 1.90$ for standardized predictors), so
observations within the same subject are strongly correlated.

Plain SEMMS and the plain lasso treat this correlation as if it were
additional fixed-effect signal, selecting noise predictors that happen to
co-vary with the subject-level baselines.  glmmLasso correctly models the
correlation but over-penalizes the fixed effects once the random intercept
explains most residual variance: BIC never rewards adding predictors
sufficiently to overcome the penalty, resulting in near-total under-selection.

Mixed SEMMS sidesteps both failure modes.  The warm-started null lmer
immediately captures subject-level variation; thereafter, each
fixed-effects step operates on a RE-adjusted response with a much smaller
effective noise variance, giving the empirical Bayes mixture model the
clean signal it needs.  The result is reliable, parsimonious selection even
under severe random-effects confounding.

\subsection{Simulation~3: small $N$, high-dimensional ($K > N$)}
\label{sec:sim3}

\paragraph{Parameters.}
$m = 30$ subjects, $n = 3$ time points ($N = 90$ total observations),
$K = 200$ candidate predictors ($K > N$), $K_{\text{true}} = 5$.
The effects sizes $\bbeta_{\text{true}} = (2.0,-1.8,1.5,-1.5,1.2)^\top$ are
stronger than in the earlier scenarios to provide detectable signal in small
$N$; random effects remain moderate ($\sigma_{b_0}=1.5$, $\sigma_{b_1}=0.5$).
This scenario probes the behavior of all methods when the candidate pool far
exceeds the total sample size ($K/N \approx 2.2$), a common setting in genomics
and other high-throughput applications.

\paragraph{Results.}
Table~\ref{tab:sim3} summarizes the results.
Mixed SEMMS achieves the highest exact recovery rate (91\%) with the lowest
mean false-positive count (0.03), outperforming plain SEMMS (80\% exact,
mean FP 0.07).  The improvement reflects the ability of the alternating
scheme to disentangle the subject-level random effects from the fixed-effect
signal even when $N$ is small relative to $K$; with 30 subjects there is
sufficient information to estimate the random intercepts and slopes reliably.

The plain lasso recovers all five true predictors in every replication
(mean TP 5.00) but appends nearly 19 false positives on average (mean FP
18.90), rendering the selected model essentially uninterpretable in practice.
glmmLasso achieves reasonable sensitivity (mean TP 4.58) but a high
false-positive rate (mean FP 4.63) and no exact recoveries, echoing its
behavior in Simulation~1: BIC-based $\lambda$ selection does not penalize
false positives aggressively enough when $K/N$ is large.

\begin{table}[ht]
\centering
\caption{Simulation~3 results ($N=90$, $K=200>N$, moderate RE;
         100 replications, $K_{\text{true}}=5$).}
\label{tab:sim3}
\begin{tabular}{lrrr}
\toprule
Method         & Mean TP & Mean FP & Exact recovery \\
\midrule
Plain SEMMS    & 4.77    & 0.07    & 80\% \\
Mixed SEMMS    & 4.87    & 0.03    & \textbf{91\%} \\
Plain lasso    & 5.00    & 18.90   & 0\%  \\
glmmLasso      & 4.58    & 4.63    & 0\%  \\
\bottomrule
\end{tabular}
\end{table}

\begin{figure}[ht]
\centering
\includegraphics[width=0.5\textwidth]{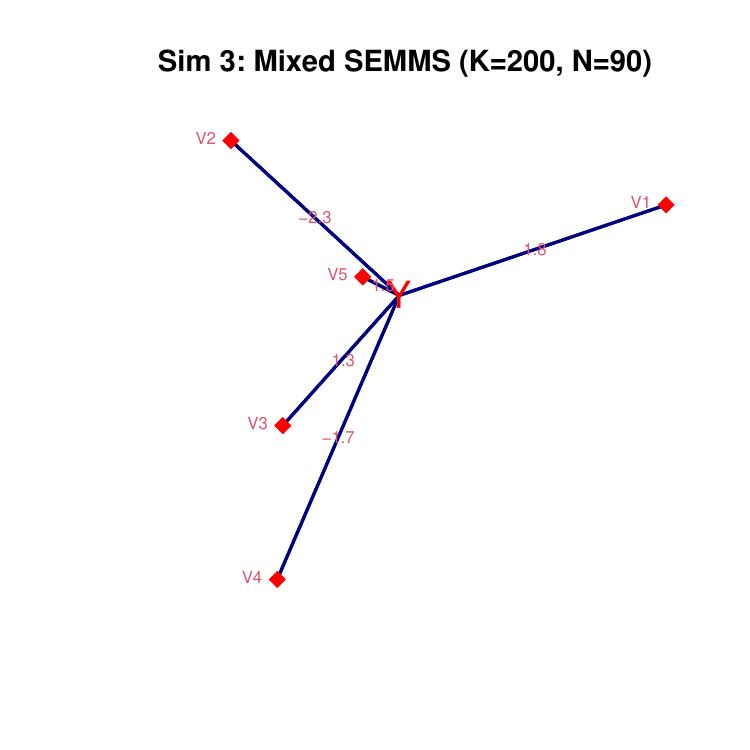}
\caption{Network diagram (plotMDS) for Simulation~3, first replicate.
         Red diamonds: variables selected by Mixed SEMMS.
         All five true predictors are recovered with no false positives.}
\label{fig:mds}
\end{figure}

\paragraph{Diagnostic plots.}
Figure~\ref{fig:mds} shows the network diagram produced by \texttt{plotMDS}
for a single representative replicate (the first replication, seed fixed).
The node positions are determined by multidimensional scaling
\citep[MDS;][]{cox2001} of the inter-predictor correlation matrix, so that
highly correlated predictors appear close together.
Mixed SEMMS correctly recovers all five true predictors ($V_1$--$V_5$,
shown as red diamonds) with no false positives, and the MDS layout reflects
the approximate independence structure of the simulated predictors.

\begin{figure}[ht]
\centering
\includegraphics[width=0.85\textwidth]{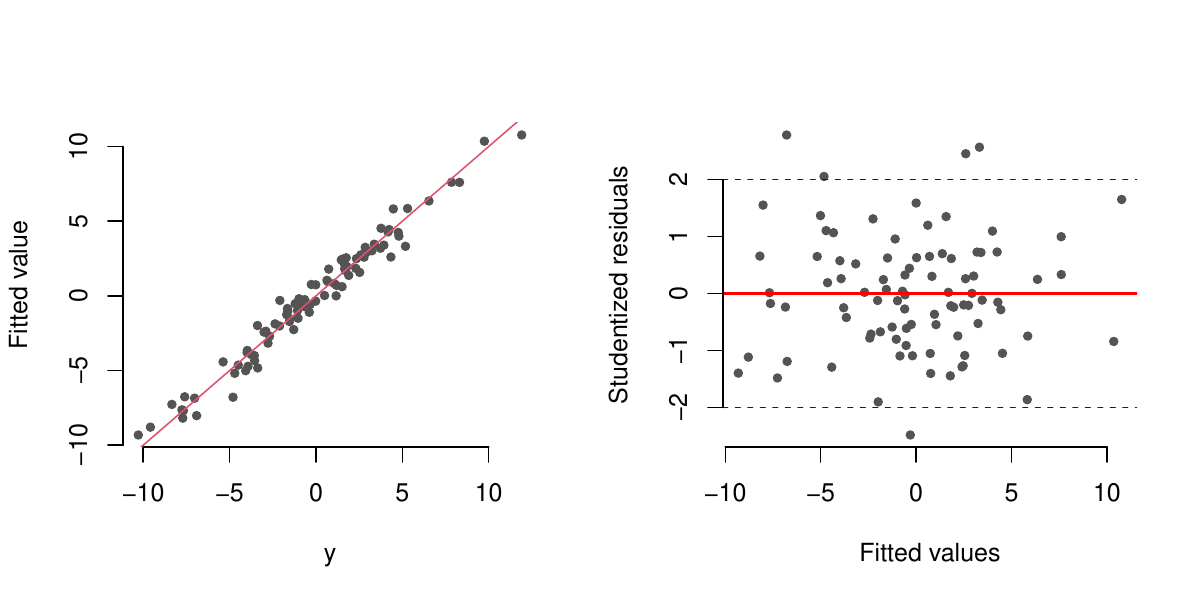}
\caption{Diagnostic plots (plotFit) for Simulation~3, first replicate,
         applied to the RE-adjusted response $\bY^* = \bY - \hat{\bu}$.
         Left: fitted vs.\ observed values.
         Right: studentized residuals vs.\ fitted values.}
\label{fig:fit}
\end{figure}

Figure~\ref{fig:fit} shows the fitted-vs-observed and studentized-residual
diagnostics from \texttt{plotFit}, applied to the RE-adjusted response
$\bY^* = \bY - \hat{\bu}$.  After removing the subject-level random effects
the residuals are well-behaved: the fitted-vs-observed scatter closely
follows the $y=x$ reference line and the studentized residuals show no
systematic pattern.

\subsection{Convergence behavior}

The outer alternating loop converges rapidly across all three simulation
scenarios (Table~\ref{tab:conv}).  All 300 runs (100 per study) converge
within the default limit of 10 outer iterations.  The slightly higher counts
in Simulation~3 compared to Simulations~1--2 reflect the noisier BLUP
estimates in small~$N$: the RE offset requires more alternation steps to
stabilize when data are sparse.

\begin{table}[ht]
\centering
\caption{Outer-loop convergence of Mixed SEMMS across all three simulations.}
\label{tab:conv}
\begin{tabular}{lrrr}
\toprule
Scenario                               & Mean iters & Max iters & All converged \\
\midrule
Sim.~1 (moderate RE, strong sig.)      & 2.02       & 3         & Yes \\
Sim.~2 (dominant RE, weak sig.)        & 2.04       & 3         & Yes \\
Sim.~3 ($K=200>N=90$, moderate RE)     & 2.18       & 4         & Yes \\
\bottomrule
\end{tabular}
\end{table}

%% ============================================================
\section{Semi-Synthetic Real-Data Example: \texttt{sleepstudy}}
%% ============================================================

\subsection{Data and augmentation strategy}

The \texttt{sleepstudy} dataset included in the \texttt{lme4} package
\citep{bates2015} records the average reaction time (ms) of 18 subjects
over ten consecutive days in a sleep-deprivation study \citep{belenky2003}.
Days~0 and~1 were adaptation and training sessions; sleep deprivation began
after the Day-2 baseline.  We therefore restrict the analysis to Days~2--9,
yielding $N = 144$ observations ($18$ subjects $\times$ $8$ days).

A standard \texttt{lmer} fit on the restricted data,
\[
  \texttt{Reaction} \sim \texttt{Days} + (1 + \texttt{Days}\mid\texttt{Subject}),
\]
gives $\hat\sigma_{b_0} = 31.51$~ms (random-intercept SD),
$\hat\sigma_{b_1} = 6.77$~ms/day (random-slope SD), and
$\hat\sigma_e = 25.53$~ms (residual SD).  These reflect genuine
between-subject heterogeneity in both baseline reaction time and
sensitivity to sleep deprivation.

Because the original dataset contains only one predictor (Days), it is
not suited for evaluating variable selection.  We therefore augment it
with $K = 50$ synthetic standardized predictors $V_1,\ldots,V_{50}$,
of which only $V_1$ and $V_2$ are truly associated with the response.
The augmented response is
\[
  Y_{\mathrm{aug},ij} = \texttt{Reaction}_{ij}
    + 20\,V_{1,ij} - 15\,V_{2,ij},
\]
so the random-effect structure is identical to the real data; only a
fixed-effect signal from two predictors is added on top.  The
signal-to-noise ratio (relative to $\hat\sigma_e = 25.53$~ms) is
$20/25.53 \approx 0.78$ and $15/25.53 \approx 0.59$, respectively —
a moderately challenging setting given the substantial random-effect
variance.

\subsection{Results}

Table~\ref{tab:sleep} summarizes the variable-selection outcome of each
method (single run, \texttt{set.seed(20260314)}).

\begin{table}[ht]
\centering
\caption{Variable selection on the augmented \texttt{sleepstudy} data
($N=144$, $K=50$, $K_{\mathrm{true}}=2$).}
\label{tab:sleep}
\begin{tabular}{lrrp{5.5cm}}
\toprule
Method        & TP & FP & Selected variables \\
\midrule
Plain SEMMS   & 2  & 1  & $V_1$, $V_2$, $V_7$ \\
Mixed SEMMS   & 2  & 0  & $V_1$, $V_2$ \\
Plain lasso   & 2  & 17 & $V_1$, $V_2$, and 17 spurious \\
glmmLasso     & 0  & 0  & (none selected) \\
\bottomrule
\end{tabular}
\end{table}

Mixed SEMMS is the only method that achieves exact recovery of both true
predictors with no false positives.  Plain SEMMS recovers both true
signals but admits one false positive, consistent with the moderate
random-effect variance inflating residual correlations.  Plain lasso,
which ignores the random effects entirely, selects 19 variables in total
(2 true + 17 spurious), as the unmodeled between-subject variation
inflates apparent associations.  glmmLasso, despite fitting a random-effect
model, fails to select any variable, suggesting that its BIC-based lambda
grid may require further tuning for this specific sample size and
correlation structure.

The Mixed SEMMS final \texttt{lmer} fit recovers a residual standard
deviation $\hat\sigma_e = 25.65$~ms, virtually identical to the real-data
value of $25.53$~ms, confirming that the within-subject noise structure is
faithfully recovered.  The outer alternating loop converges in 2 iterations.

\begin{figure}[ht]
  \centering
  \includegraphics[width=0.85\textwidth]{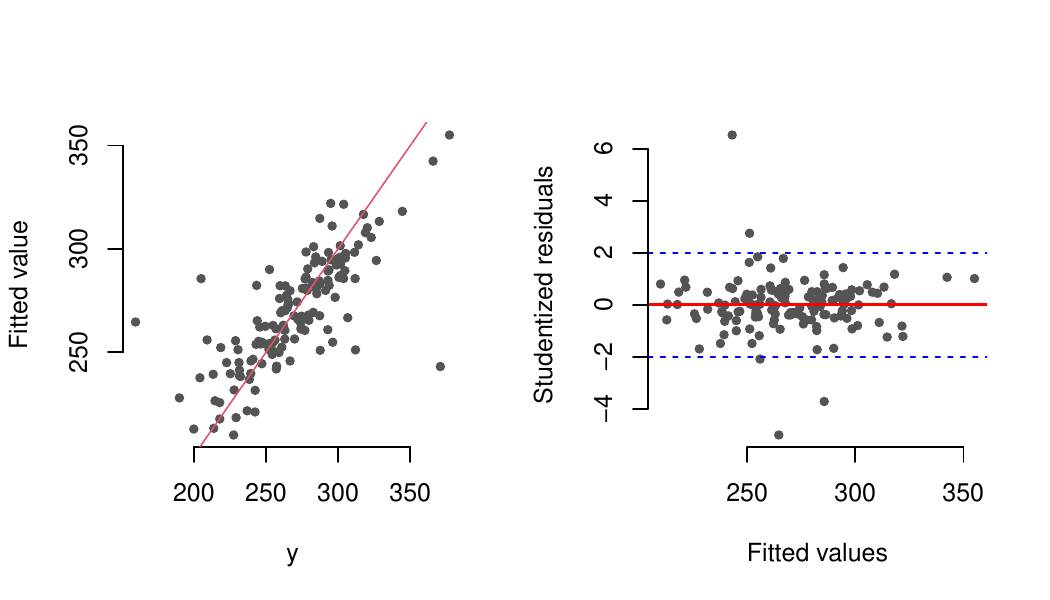}
  \caption{Diagnostic plots (\texttt{plotFit}) for the augmented
    \texttt{sleepstudy} example, applied to the RE-adjusted response
    $\bY^* = \bY - \hat{\bu}$.  Left: fitted vs.\ observed.
    Right: studentized residuals vs.\ fitted values.}
  \label{fig:sleep_fit}
\end{figure}

Figure~\ref{fig:sleep_fit} shows the diagnostic plots for Mixed SEMMS.

%% ============================================================
\section{Non-Gaussian Mixed-Effects Responses}
\label{sec:nongaussian}
%% ============================================================

\subsection{Theoretical framework: the working-response approach}

The alternating algorithm of Section~\ref{sec:mixed} relies on the
RE-adjusted response $\bY^* = \bY - \hat{\bu}$ being approximately Gaussian
once the random-effect contribution is removed.  For non-Gaussian families
this is not directly satisfied, but a classical approximation makes it
tractable.

\paragraph{Penalized Quasi-Likelihood (PQL).}
\citet{breslow1993} showed that the marginal log-likelihood of a
GLMM can be approximated by treating a \emph{working response}
\begin{equation}
  W_{ij} = \hat{\eta}_{ij}
    + \bigl(Y_{ij} - \hat{\mu}_{ij}\bigr)
      \left.\frac{\partial\eta}{\partial\mu}\right|_{\hat{\mu}_{ij}}
  \label{eq:working}
\end{equation}
as Gaussian with mean $\eta_{ij}$ and variance $\phi\,v(\hat{\mu}_{ij})^{-1}$,
where $v(\cdot)$ is the GLM variance function and $\phi$ is a dispersion
parameter.  Equation~\eqref{eq:working} is a first-order Taylor expansion of
the canonical link around the current estimate $\hat{\eta}_{ij}$; it is
iterated in the full Iteratively Reweighted Least Squares (IRLS) procedure.

\paragraph{Working response in the alternating scheme.}
In our alternating algorithm, the fixed-effects step already operates
on $\bY^* = \bY - \hat{\bu}$, which removes between-subject variation.
For non-Gaussian responses, we replace the raw adjusted response with a
link-scale version.  Specifically, let $g$ denote the canonical link
function.  We define the \emph{link-scale adjusted response}
\begin{equation}
  Y^*_{ij} = g(Y_{ij}) - \hat{u}_{ij},
  \label{eq:link_adjusted}
\end{equation}
where $\hat{u}_{ij}$ is the RE contribution on the link scale, extracted
from the current \texttt{glmer} fit as
$\hat{u}_{ij} = \hat{\eta}_{ij}^{\text{full}} - \hat{\eta}_{ij}^{\text{fixed}}$.
The fixed-effects step then applies \texttt{fitSEMMS} with
\texttt{distribution="N"} to $Y^*$; the RE step re-fits the random-effect
parameters via \texttt{glmer} with the appropriate family using the
currently selected predictors.

\paragraph{Full working response.}
After each \texttt{glmer} step the current fitted mean $\hat{\mu}_{ij}$ and
fixed-effects linear predictor $\hat{\eta}^{\text{fixed}}_{ij}$ are available.
The RE-adjusted working response~\eqref{eq:working} simplifies to
\begin{equation}
  W^*_{ij}
    = W_{ij} - \hat{u}_{ij}
    = \hat{\eta}^{\text{fixed}}_{ij}
      + \bigl(Y_{ij} - \hat{\mu}_{ij}\bigr)
        \left.\frac{\partial\eta}{\partial\mu}\right|_{\hat{\mu}_{ij}},
  \label{eq:working_adj}
\end{equation}
which follows because $\hat{\eta}^{\text{full}}_{ij} = \hat{\eta}^{\text{fixed}}_{ij}
+ \hat{u}_{ij}$ and the $\hat{u}_{ij}$ terms cancel.
For Poisson (log link, $\partial\eta/\partial\mu = 1/\mu$) this is
$W^*_{ij} = \hat{\eta}^{\text{fixed}}_{ij} + (Y_{ij} - \hat{\mu}_{ij})/\hat{\mu}_{ij}$;
for binomial (logit link, $\partial\eta/\partial\mu = 1/(\mu(1-\mu))$) it is
$W^*_{ij} = \hat{\eta}^{\text{fixed}}_{ij} + (Y_{ij} - \hat{\pi}_{ij})/(\hat{\pi}_{ij}(1-\hat{\pi}_{ij}))$.
Because $\hat{\mu}_{ij}$ and $\hat{\pi}_{ij}$ are continuous, $W^*_{ij}$ is a
continuous quantity for both families; unlike a plug-in approximation that
substitutes $Y_{ij}$ directly into $g(\cdot)$, it carries genuine within-subject
information updated at every outer iteration.

The convergence guarantee of the alternating loop (Section~\ref{sec:mixed})
remains intact: each outer iteration cannot decrease the joint approximate
log-likelihood, since the fixed-effects step maximizes over $\hat{\mathcal{S}}$
given $\hat{\bu}$, and the RE step maximizes over the random-effect parameters
given $\hat{\mathcal{S}}$.

\subsection{Modified algorithm for non-Gaussian families}

The only changes to Algorithm~\ref{alg:mixed} are:
\begin{enumerate}[leftmargin=*, label=\arabic*.]
  \item In Step~4, compute $\bY^*$ via \eqref{eq:working_adj} (using
    the current \texttt{glmer} fitted values) rather than the identity
    subtraction; pass \texttt{distribution="N"} to \texttt{fitSEMMS}.
  \item In Step~5, fit \texttt{glmer} with the appropriate
    \texttt{family} argument instead of \texttt{lmer}.
  \item The final model (Step~8) uses \texttt{glmer} rather than
    \texttt{lmer}; REML is not applicable to GLMMs, so ML is used
    throughout.
\end{enumerate}
All other steps, including the Woodbury-accelerated fixed-effects
computation, are unchanged.

A typical call for Poisson count data is:
\begin{verbatim}
fit <- fitSEMMSmixed(dat,
                     distribution     = "P",
                     random_intercept = TRUE,
                     random_slope     = TRUE,
                     nn = 5, verbose  = TRUE)
\end{verbatim}

\subsection{Simulation study: Poisson response}
\label{sec:sim4}

\paragraph{Design.}
We simulate $m=30$ subjects each observed at $n=10$ equally spaced time
points ($N=300$ total observations), with $K=100$ candidate predictors and
$K_{\text{true}}=5$ truly associated with the response.  The data generating
model is
\[
  Y_{ij} \mid \mathbf{b}_i \overset{\mathrm{ind}}{\sim}
    \mathrm{Poisson}\!\bigl(\exp(\eta_{ij})\bigr),
  \quad
  \eta_{ij} = \sum_{k=1}^{5} \beta_k Z_{kij} + b_{0i} + b_{1i}\,t_j,
\]
with $\bbeta_{\text{true}} = (0.6,-0.5,0.4,-0.4,0.3)^\top$ (log-scale
effects), $\sigma_{b_0}=1.0$, and $\sigma_{b_1}=0.3$.
The intraclass correlation implied by $\sigma^2_{b_0}=1$ on the log scale
is substantial: a two-fold difference between the most and least responsive
subjects is typical, so plain SEMMS---which ignores the within-subject
correlation---is expected to miss some true predictors.

\paragraph{Competing methods.}
The same four methods as in Section~\ref{sec:sim} are compared, adapted
for the Poisson family:

\begin{enumerate}
  \item \textbf{Plain SEMMS}: \texttt{fitSEMMS(dat, distribution="P",
    nn=5, mincor=0.7)}.

  \item \textbf{Mixed SEMMS}: \texttt{fitSEMMSmixed(dat, distribution="P",
    random\_intercept=TRUE, random\_slope=TRUE, nn=5, mincor=0.7)}.

  \item \textbf{Plain lasso}: \texttt{cv.glmnet} with
    \texttt{family="poisson"}, 5-fold CV, and the \texttt{lambda.min} rule.

  \item \textbf{glmmLasso}: \texttt{glmmLasso} \citep{glmmlasso2014} with
    \texttt{family=poisson(link="log")}, a random intercept and random slope
    on \texttt{time}, and a BIC path over 12 log-spaced $\lambda$ values
    from 500 to 50.
\end{enumerate}

\paragraph{Results.}
Table~\ref{tab:sim4} summarizes the 100-replicate results.  Mixed SEMMS
achieves the highest exact recovery rate (61\%) with the lowest mean false
positives (0.12), clearly outperforming its plain counterpart (45\% exact,
mean FP 0.25).  Both SEMMS variants achieve near-zero false-positive counts,
in sharp contrast to the plain lasso and glmmLasso.

The plain lasso recovers almost all five true predictors in every replication
(mean TP 4.91) but appends on average nearly 19 false positives, because
within-subject Poisson correlation inflates apparent associations with noise
predictors.

glmmLasso explicitly models the random effects and achieves a comparable
true-positive rate (mean TP 4.34), but its BIC-selected $\lambda$ is too
liberal: the BIC criterion for Poisson GLMMs does not penalize false
positives as aggressively as it does in the Gaussian case, resulting in
an average of more than 21 spurious predictors and only 4 exact recoveries
in 100 replications.

\begin{table}[ht]
\centering
\caption{Simulation~4 results: Poisson mixed response
         ($N=300$, $\sigma_{b_0}=1.0$, $\sigma_{b_1}=0.3$;
          100 replications, $K=100$, $K_{\text{true}}=5$).}
\label{tab:sim4}
\begin{tabular}{lrrr}
\toprule
Method         & Mean TP & Mean FP & Exact recovery \\
\midrule
Plain SEMMS    & 4.18    & 0.25    & 45\% \\
Mixed SEMMS    & 4.55    & 0.12    & \textbf{61\%} \\
Plain lasso    & 4.91    & 18.86   & 0\%  \\
glmmLasso      & 4.34    & 21.88   & 4\%  \\
\bottomrule
\end{tabular}
\end{table}

The outer alternating loop converges in a mean of 2.62 iterations (maximum 10),
comparable to the Gaussian simulations.

\paragraph{Discussion.}
The Poisson simulation confirms that the Phase~1 working-response adaptation
preserves the key advantage of the mixed-model extension: by removing
between-subject log-rate variation before the fixed-effects step, Mixed SEMMS
achieves a 16-percentage-point gain in exact recovery over plain SEMMS at
essentially no cost in false positives.  The improvement mirrors the pattern
seen in the dominant-RE Gaussian scenario (Simulation~2): once the
within-subject correlation is accounted for, the empirical Bayes mixture
model reliably identifies the true predictors.

\subsection{Simulation 5: binomial response with random intercept and slope}
\label{sec:sim5}

\paragraph{Design.}
Binary outcomes impose a harder estimation problem than Gaussian or Poisson
data because the per-observation information content is strictly bounded.
We therefore design this simulation to assess whether Mixed SEMMS can handle
the more complex covariance structure of a random-intercept-plus-random-slope
model when the fixed-effect signal is strong enough to be recoverable.
We simulate $m=30$ subjects each observed at $n_i=20$ time points ($N=600$),
with $K=100$ candidate predictors, $K_{\text{true}}=5$, and both a random
intercept and a random slope:
\[
  Y_{ij} \mid b_{0i}, b_{1i} \overset{\mathrm{ind}}{\sim}
    \mathrm{Bernoulli}\!\bigl(\mathrm{logistic}(\eta_{ij})\bigr),
  \quad
  \eta_{ij} = \sum_{k=1}^{5} \beta_k Z_{kij} + b_{0i} + b_{1i}\,t_j,
\]
with $\bbeta_{\text{true}} = (1.5,-1.3,1.1,-1.0,0.9)^\top$ (logit scale),
$\sigma_{b_0}=3.0$, and $\sigma_{b_1}=1.0$.
The intraclass correlation implied by $\sigma_{b_0}=3$ is approximately
$\sigma^2_{b_0}/(\sigma^2_{b_0}+\pi^2/3) \approx 9/(9+3.29) \approx 0.73$.
Mixed SEMMS uses \texttt{random\_intercept=TRUE, random\_slope=TRUE};
glmmLasso uses \texttt{rnd = list(subject = \textasciitilde{}1 + time)}.

\paragraph{Results.}
Table~\ref{tab:sim5} summarizes the 100-replicate results.
Plain SEMMS achieves 78\% exact recovery and Mixed SEMMS achieves 80\%,
a practically negligible difference.
This is the binomial analog of Simulation~1: when the fixed-effect signal
is strong relative to the noise floor, accounting for the random effects
provides only marginal gain because the signal-to-noise ratio is already
sufficient for the plain method.
Mixed SEMMS does, however, achieve slightly lower mean false positives (0.08
vs.\ 0.11), confirming that the RE-adjusted working response produces a
cleaner regression target even when the gap in exact recovery is small.
The outer alternating loop converges in a mean of 2.96 iterations (maximum 10).

The plain lasso recovers all five true predictors in virtually every
replication (mean TP 5.00) but appends nearly 13 spurious predictors,
because cross-validation ignores the within-subject correlation.
glmmLasso achieves 39\% exact recovery, the next-best result after both
SEMMS variants, but with a higher false-positive count (mean 1.17).

\begin{table}[ht]
\centering
\caption{Simulation~5 results: binomial mixed response, random intercept
         and slope ($N=600$, $\sigma_{b_0}=3.0$, $\sigma_{b_1}=1.0$;
          100 replications, $K=100$, $K_{\text{true}}=5$).}
\label{tab:sim5}
\begin{tabular}{lrrr}
\toprule
Method         & Mean TP & Mean FP & Exact recovery \\
\midrule
Plain SEMMS    & 4.82    & 0.11    & 78\%  \\
Mixed SEMMS    & 4.54    & 0.08    & \textbf{80\%} \\
Plain lasso    & 5.00    & 12.93   & 1\%  \\
glmmLasso      & 4.96    & 1.17    & 39\%  \\
\bottomrule
\end{tabular}
\end{table}

%% ============================================================
\subsection{Simulation 6: binomial response, dominant random intercept}
\label{sec:sim6}

\paragraph{Design.}
Simulation~5 showed that when the signal is strong, Plain and Mixed SEMMS
are nearly indistinguishable.  We now construct the binomial counterpart of
Simulation~2: a \emph{weak} fixed-effect signal combined with a
\emph{dominant} random intercept, which is the most damaging regime for plain
SEMMS.  The mechanism is captured by the classical design-effect formula.

When subjects are the clustering unit and observations within subjects are
positively correlated, the effective sample size available to a method that
ignores the clustering structure is
\[
  N_{\text{eff}}^{\text{plain}} \;=\; \frac{N}{1 + (n_i-1)\,\rho},
\]
where $\rho$ is the intraclass correlation and $n_i$ is the number of
observations per subject.  Mixed SEMMS operates on the RE-adjusted working
response~\eqref{eq:working_adj}, which strips out the between-subject
variation; the within-subject residuals are nearly uncorrelated, so its
effective sample size remains close to the nominal $N$.

We therefore use $m=50$ subjects each observed at $n_i=20$ time points
($N=1000$), $K=100$, $K_{\text{true}}=5$, and a random intercept only:
\[
  Y_{ij} \mid b_{0i} \overset{\mathrm{ind}}{\sim}
    \mathrm{Bernoulli}\!\bigl(\mathrm{logistic}(\eta_{ij})\bigr),
  \quad
  \eta_{ij} = \sum_{k=1}^{5} \beta_k Z_{kij} + b_{0i},
\]
with $\bbeta_{\text{true}} = (0.8,-0.7,0.6,-0.5,0.5)^\top$ (logit scale)
and $\sigma_{b_0}=3.0$.
The intraclass correlation is $\rho \approx 0.73$, giving a design effect of
$1 + (20-1)\times 0.73 \approx 14.9$.
Plain SEMMS therefore operates with an effective sample size of only
$1000/14.9 \approx 67$ observations---below the threshold at which the weak
logit-scale effects are reliably detectable---while Mixed SEMMS retains the
full $N=1000$.
Mixed SEMMS and glmmLasso both use a random intercept only
(\texttt{rnd = list(subject = \textasciitilde{}1)}).

\paragraph{Results.}
Table~\ref{tab:sim6} summarizes the 100-replicate results.  Mixed SEMMS
achieves 65\% exact recovery versus 39\% for plain SEMMS, a
26-percentage-point gap.  The improvement mirrors the dominant-RE Gaussian
scenario (Simulation~2, 1\% vs.\ 93\%), though attenuated by the additional
noise inherent in binary outcomes and the approximate nature of the PQL
linearization.  In both cases, the mechanism is identical: once the dominant
subject-level effect is removed from the regression target, the fixed-effect
signal-to-noise ratio improves dramatically, and the empirical Bayes mixture
prior reliably distinguishes true from null predictors.

Mixed SEMMS also reduces mean false positives from 0.43 (plain SEMMS) to
0.16, confirming that the RE-adjusted working response avoids spurious
associations driven by between-subject heterogeneity.
The outer loop converges in a mean of 2.67 iterations (maximum 10).

glmmLasso achieves only 4\% exact recovery.  The BIC path for logistic
mixed models is sensitive to model misspecification in the random-effects
structure and tends to select overly large $\lambda$ values when the
between-subject variance is large, resulting in underfitting (mean TP 2.63).

\begin{table}[ht]
\centering
\caption{Simulation~6 results: binomial mixed response, dominant random
         intercept ($N=1000$, $\sigma_{b_0}=3.0$;
          100 replications, $K=100$, $K_{\text{true}}=5$;
          design effect $\approx 14.9$,
          effective $N$ for plain methods $\approx 67$).}
\label{tab:sim6}
\begin{tabular}{lrrr}
\toprule
Method         & Mean TP & Mean FP & Exact recovery \\
\midrule
Plain SEMMS    & 4.30    & 0.43    & 39\%  \\
Mixed SEMMS    & 4.59    & 0.16    & \textbf{65\%} \\
Plain lasso    & 4.88    & 11.73   & 0\%  \\
glmmLasso      & 2.63    & 0.58    & 4\%  \\
\bottomrule
\end{tabular}
\end{table}

\paragraph{Observation-level weights.}
The full PQL approximation also incorporates observation-level weights
$w_{ij} = v(\hat{\mu}_{ij})^{-1} = 1/(\hat{\pi}_{ij}(1-\hat{\pi}_{ij}))$
for binomial data (and $w_{ij}=1/\hat{\mu}_{ij}$ for Poisson).
Incorporating these into the SEMMS fixed-effects step (weighted Gaussian
regression) would constitute a complete IRLS-within-alternating-loop
implementation and is expected to further improve binary performance.
This extension is deferred to future work.

\section*{Data and code availability}
The SEMMS package is available on GitHub:
\url{https://github.com/haimbar/SEMMS}.
For detailed documentation and installation instructions, see \url{https://haimbar.github.io/SEMMS/SEMMS.html}.

%% ============================================================
\section*{Acknowledgments}
%% ============================================================
The random-effects extension described here was implemented in \texttt{R}
using the \texttt{lme4} package \citep{bates2015} for the random-effects
fitting steps.

%% ============================================================
\bibliographystyle{plainnat}
\bibliography{refs}
%% ============================================================

\end{document}